\documentclass[12pt,a4paper]{article}       

\begin{document}

\renewcommand{\thesection}{\arabic{section}}

\setcounter{equation}{0}
\setcounter{section}{0}

\renewcommand{\thefootnote}{\alph{footnote}}

{\vskip -0.5in\hfill\hfil{\rm\normalsize Printed on \today}
\vskip 0.4in}

{\centerline{\Large \bf {Sigma Model Lagrangian for the  
Heisenberg Group}}}

\vskip 1.1cm

{\center 

Belal E. Baaquie\footnote{E-mail: uspbeb@nus.edu.sg} \\
{\it Department of Physics, National University of Singapore, 
Singapore 119260} \\

\vskip 0.5cm

and \\ 

\vskip 0.5cm
   
Kok Kean Yim\footnote{E-mail: yimkk@pacific.net.sg} \\
{\it Department of Physics, National University of Singapore,
Singapore 119260} 
\\

}

\

\

\

\

\begin{abstract}

\ 

We study the Lagrangian for a sigma model based on the non-compact Heisenberg
group. A unique feature of this model -- unlike the case for compact Lie groups -- is that the definition of the Lagrangian has to be regulated since the trace over the Heisenberg group is otherwise divergent. The resulting theory is a real Lagrangian with a quartic interaction term. After a few non-trivial transformations, the Lagrangian is shown to be equivalent -- at the classical level -- to a complex cubic Lagrangian. A one loop computation shows that the quartic and cubic Lagrangians are equivalent at the quantum level as well.

The complex Lagrangian is known to classically equivalent to the $SU(2)$ sigma model, with the equivalence breaking down at the quantum level. An explanation of this well known results emerges from the properties of the Heisenberg sigma model.

\end{abstract}
{\centerline{PACS: 00.00.Aa, 00.00.Aa, 00.00.Aa, 00.00.Aa.}} 
\newpage
\setcounter{equation}{0}
\section{\normalsize \bf Introduction}

\
\indent 

Sigma models in two-dimensional spacetime are ubiquitous in theoretical physics, with many applications in quantum field theory. Sigma models are usually based on compact Lie groups.

In this paper, we construct a sigma model based on the non-compact Heisenberg group. The present study is motivated by the more complex case studied in \cite{baaquie7} where a supersymmetric Yang-Mills theory was obtained having a local infinite dimensional Kac-Moody group as its gauge 
group. The need for regulating the Lagrangian of the theory was essential in obtaining local Kac-Moody gauge symmetry. The Heisenberg algebra in an infinite dimensional non-compact subalgebra of the Kac-Moody algebra, the Heisenberg sigma model is the simplest theory having the new features that emerge from constructing quantum field theories based on infinite dimensional Lie algebras. This is the main motivation for studying the Lagrangian obtained in this article.  

Given that the group element of the Heisenberg group is infinite dimensional, the usual procedure of obtaining a lagrangian by tracing over a representation of the group yields a divergent result. We regulate the trace to obtain a finite Lagrangian, which turns out to be a real quartic lagragian $\mathcal{L}_4$. After some straightforward calculation, the quartic is shown to be equivalent an imaginary cubic Lagrangian $\mathcal{L}_3$. Interestingly enough, the relationship  of the two Lagrangians is {\bf not} that of a duality transformation since the mapping does not induce an inversion of the coupling constant.

The cubic Lagrangian $\mathcal{L}_3$ in turn is known to be classically equivalent to the sigma model based on the $SU(2)$ group \cite{cz}; and furhermore, it is also known that this equivalence breaks down on quantizing the two theories \cite{nappi}. Our derivation provides an understanding of this quantum inequivalence, since a quantum theory based on a compact group is unlikely to have the same properties under renormalization to a similar classical theory based on a non-compact group. 

In this article we will employ the backgound field method to study the renormalizability of the theory  up to $1$-loop correction. A similar, but much more complex,  calculation was carried out in \cite{br} to study the renormalizability of a $U(1)$ gauge field with Kac-Moody gauge symmerty. We will calculate the  $\beta$-function for both the quartic and cubic Lagrangian realizations of the theory, and show that to one loop they are identical. We will hence establish the one-loop quantum equivalence of two apparently dissimilar bosonic theories. 

\section{\normalsize \bf The Heisenberg Sigma Model Lagrangian}

Consider the (non-compact) Heisenberg Algebra
\begin{eqnarray}
[ \, x \, , \, p \, ] = i k
\end{eqnarray}
which, in terms of the creation and destruction operators, is
given by 
\begin{eqnarray}
[ \, a \, , \, a^{\dagger} \, ] = k \, .
\end{eqnarray}
Since we would like to construct a sigma model based on the Heisenberg algebra, and we start from the finite group elements of the Heisenberg algebra. By the usual exponential mapping, we can write
\begin{eqnarray}
\Omega & = & \mbox{exp} \mbox{\large[} i \phi + i \omega a
+ i {\omega}^{*} a^{\dagger} \mbox{\large]} \, 
\end{eqnarray}

Note that the field (group coordinate) $ \phi $ is a real 
variable, whereas $ \omega $ is an arbitrary complex variable. 
The field $ \phi $  has to be introduced due to 
the existence of the central extension of the algebra. 

The simplest nonlinear sigma model Lagrangian that gives a space-time
dependence to the group coordinates $ \phi $ and $ \omega $ is 
defined by
\begin{eqnarray}
\mathcal{L} &=& \mbox{Tr} \mbox{\large[} \partial_{\mu} \Omega^{\dagger} 
\partial_{\mu} \Omega \mbox{\large]} \, . \label{lag0}
\end{eqnarray}
However, this approach fails since the trace over the non-compact 
operators $ a $, $ a^{\dagger} $ diverges, yielding 
\begin{eqnarray}
\mathcal{L} &=& \infty \, .
\end{eqnarray}  
A similar situation was encountered in a more complex setting of
defining supersymmetric gauge fields with the infinite dimensional Kac-Moody symmetry
\cite{baaquie7}.

To successfully obtain a finite Lagrangian, one must  
regularize the trace $ \mbox{Tr} [ \, \cdots \, ] $ over the 
infinite dimensional operators. There is a wide variety of 
regulators which one can choose, and we expect from the principle of universality that a whole range of regulators would lead to the same renormalizable theory \cite{baaquie7}. 
We make the natural choice for the regulator of the trace given by
\begin{eqnarray}
e^{ - \rho a^{\dagger} a} ~:~\mathrm{Trace~Regulator}
\label{rf}
\end{eqnarray}
In direct analogy with the procedure given in  \cite{baaquie7}, we define a generalization of the non-linear sigma model by the following 
\begin{eqnarray}
\label{lww}
\mathcal{L} &=& { 1 \over 2  \lambda } \mbox{Tr} 
\mbox{\large[} e^{ - \rho a^{\dagger} a } \partial_{\mu} 
\Omega^{\dagger} 
\partial_{\mu} \Omega \mbox{\large]} \label{lag1}
\end{eqnarray}
where a normalization constant of $ (2 \lambda )^{-1} $ has been 
included. We have added the regularization factor  $ e^{ - \rho a^{\dagger} a } $ in the Lagrangian so that the trace over the infinite dimensional $a$ and $a^{\dagger}$
converges. Thus, we have a Lagrangian that resembles the Lagrangian of the $U(N)$ chiral model  
or non-linear sigma model, and that is valid for the infinite dimensional Heisenberg group. 

To write the Lagrangian in terms of the real fields, $ \alpha $ and $ \beta $, define
\begin{eqnarray}
\omega (x) &=& \alpha (x) + i \beta (x)
\end{eqnarray}
We have
\begin{eqnarray}
\label{ww}
-i\Omega^\dagger \partial_\mu \Omega = \partial_\mu \phi + k(\beta \partial_\mu \alpha - \alpha \partial_\mu \beta) +\partial_\mu\alpha a + \partial_\mu\beta a^\dagger
\end{eqnarray}

Choosing the constant factor $ \lambda $ such that
\begin{eqnarray}
\lambda &=& \mbox{Tr} \mbox{\large[} e^{ - \rho a^{\dagger} a } 
\mbox{\large]}
\end{eqnarray}
the Lagrangian (\ref{lag1}) yields, using eq.(\ref{ww}), the following
\begin{eqnarray}
\mathcal{L}
&=& {1 \over 2 } \mbox{\large[} \, \{\partial_{\mu} \phi - k( \beta \partial_{\mu} \alpha - \alpha \partial_{\mu} 
\beta )\}^2 + f( \rho ) (\partial_{\mu} \alpha \partial_{\mu} \alpha + 
\partial_{\mu} \beta \partial_{\mu} \beta) \, \mbox{\large]} 
\label{LLL04} 
\end{eqnarray}
where 
\begin{eqnarray}
f( \rho ) 
&=& \mbox{Tr} \mbox{\large[} \mbox{\large(} a^\dagger a + 
aa^\dagger \mbox{\large)}
e^{- \rho a^\dagger a } \mbox{\large]} /
\mbox{Tr} \mbox{\large[} e^{- \rho a^\dagger a } 
\mbox{\large]} \nonumber \\
&=& \frac{k}{ \tanh (\frac{k\rho}{2})}
\end{eqnarray}
 
The Lagrangian given above describes a non-trivial quartic scalar field theory consisting of three massless scalar fields. The kinetic terms of the two fields $ \alpha (x) $, $ \beta (x) $
are scaled by the regularization constant $ \rho $ (through the function $ f ( \rho ) $) with respect to the field $\phi (x)$. We call this the quartic Lagrangian $ \mathcal{L}_4 $. 

Define the (bounded) coupling constant by
\begin{eqnarray}
g =\tanh (\frac{k\rho}{2})\in  {[ \, 0 \, , \, 1 \, ]}
\end{eqnarray}

Rescale fields by
\begin{eqnarray}
\phi (x) &\rightarrow& { \phi (x) \over g } \nonumber \\
\alpha (x) &=& [ f ( \rho ) ]^{ 1 \over 2} \, \alpha^{1} (x) \\
\beta (x) &=& [ f ( \rho ) ]^{ 1 \over 2} \, \alpha^{2} (x) 
\nonumber 
\end{eqnarray}
We then obtain the Lagrangian in the following compact form 
\begin{eqnarray}
\mathcal{L}_4 &=& { 1 \over 2 g^2 } \mbox{\large[} \partial_{\mu} 
\alpha^i \partial_{\mu} \alpha^i + \mbox{\large(} \partial_{\mu}
\phi - \epsilon_{ij} \alpha^i \partial_{\mu} \alpha^j
\mbox{\large)}^2 \mbox{\large]} \, . 
\label{lag4}
\end{eqnarray}
with the action given by
\begin{eqnarray}
S_4&=&\int dxdt \mathcal{L}_4 \\
		&=& S_{\alpha}+S_{\mathrm {quartic}}
\end{eqnarray}

Note that, if we had started with the following regulated 
Lagrangian 
\begin{eqnarray}
\mathcal{L} &=& { 1 \over 2 \lambda } \mbox{Tr}
\mbox{\large[} e^{ - \rho a^{\dagger} a } \partial_{\mu} \Omega  
\partial_{\mu} \Omega^{\dagger} \mbox{\large]}
\end{eqnarray}
instead of (\ref{lag1}), then we would get a Lagrangian identical to (\ref{LLL04}) 
since the change in ordering of the $\Omega$ field switches $g$ to $-g$. 

Dimensional analysis tells us that the coupling constant $g$ is 
dimensionless in $ d = 2 $. When one quantizes the theory in
$ d = 2 $, one can expect the theory to be renormalizable. Since the coupling constant in $d=2$ is dimensionless and bounded, at least in case of the regulator that we have used, it is possible that the renormalized coupling constant also turns out to be bounded. Thus, if this theory needs renormalization (which it does), then
as we let the coupling constant run, an educated guess is that it  should hit a fixed
point. On performig a 1-loop correction calculation for the beta function, we will see that the coupling increases for short distances; hence, we expect that the theory should have a non-trivial ultra-violet fixed point.

The quantum field theory is defined by the Feynman path integral

\begin{eqnarray}
Z=\int D\alpha_i D\phi e^{-S_4}
\end{eqnarray}

Since the field $ \phi $ appears only quadratically 
in the Lagrangian $\mathcal{L}_4$, one can integrate it  out in the
path integral to get a simpler expression. We perform the $\phi$ integration in the following manner.
\begin{eqnarray}
Z&=&\int D\alpha_i D\phi e^{-S_4}\\
&=&\int D\alpha_i e^{-S_{\alpha}} \int D\phi DA_i \exp\Big (-\frac{1}{2}\int A^2_\mu 
           +\frac{i}{g }\int A_\mu \big( \partial_{\mu}
\phi - \epsilon_{ij} \alpha^i \partial_{\mu} \alpha^j \big )\Big )
\end{eqnarray}
Performing the $\phi$ integrations yields 
\begin{eqnarray}
\partial_{\mu}A_{\mu}&=&0 \\
\Rightarrow A_{\mu}&=&\epsilon_{\mu \nu} \partial_\nu \chi
\end{eqnarray}
 
Performing the $A_\mu$  Gaussian path integrations in $ d = 2 $
space-time, we can rewrite the Lagrangian $ \mathcal{L} $ as
\begin{eqnarray}
\mathcal{L}_3 &=&   {1 \over 2 } \mbox{\large{[}} 
\partial_{\mu} \phi^{i}  
\partial_{\mu} \phi^{i}  - i { 2 \over 3} g \epsilon_{ijk} 
\epsilon_{\mu \nu} \phi^{i} 
\partial_{\mu} \phi^{j} \partial_{\nu} \phi^{k} 
\mbox{\large{]}} \qquad \label{lag3} 
\end{eqnarray}
with the notations
\begin{eqnarray}
 \alpha^1 &=& g \phi^{1}\nonumber \\
 \alpha^2 &=& g \phi^{2}\nonumber \\
 \chi &=& \phi^3 \nonumber
\end{eqnarray}
where $ \epsilon_{ijk} $ is the standard totally antisymmetric 
tensor. This form of the Lagrangian is complex with the interaction given by a single imaginary cubic interaction term. 

The cubic Lagrangian that we have obtained is identical to that of the scalar-field 
theory model that is known to be classically equivalent \cite{cz} to the $SU(2)$ sigma model; this model has spontaneous particle production\cite{nappi}.  

Recall that the $SU(2)$ sigma model in invariant under global left and right group muliplication; however, it is known that {\bf only one} of the symmetries of the $SU(2)$ sigma model is present in the (classically) equivalent cubic Lagrangian \cite{cz}. A simple explanation of this fact follows from the derivation of the cubic Lagrangian from the Heisenberg group: due to the regulator, the Lagrangian given eq.(\ref{lww}) is invariant only under the following global transformation
\begin{eqnarray}
\label{rinv}
\Omega(t,x) \rightarrow  \Phi \Omega(t,x) 
\end{eqnarray}
where $\Phi$ is a constant element of the Heisenberg group. Note multiplication on the right by
$\Phi$ is not a symmetry due to the regulator, and hence explaining the lack of invariance of the cubic Lagrangian under both left and right group muliplication.

The cubic Lagrangian also corresponds to the 
first two non-trival terms of the $SU(2)$ WZW model. 

Note that the cubic Lagrangian is almost identical to a 
Lagrangian which describes the low-energy excitation of the
one dimensional quantum antiferromagnet with short-range 
N$\acute{e}$el order \cite{fradkin1}. The difference being that 
in the antiferromagnet case the fields must reside on a sphere 
while in our case we have no such restriction.  

In particular, we note the following.

\begin{enumerate}
\item In the original Lagrangian (\ref{lag1}), the only symmetry 
that is manifest is the invariance of the Lagrangian under left multiplicative
of $ \Omega $.  Although not too obvious, this symmetry is present in the 
quartic Lagrangian $\mathcal{L}_4$. However, this original 
symmetry is hidden in the cubic Lagrangian $\mathcal{L}_3$. On the other hand, in the cubic Lagrangian $\mathcal{L}_3$ the classical symmetries of the 
sigma model are manifest, and these in turn are only implicit in the $\mathcal{L}_4$. 
\item  We have transformed a Lagrangian with quartic interaction to a
Lagrangian with a cubic interaction using Gaussian integrations.
\item The coupling constant $g$ is identically transformed from
one Lagrangian into one another. This contrasts  with the 
usual duality transformation which inverts the strength of $g$. 
\item Since the interaction in the cubic Lagrangian is one order 
lower than the quartic Lagrangian (\ref{lag4}), one can expect 
its renormalization to be relatively simpler. The limitation 
of the cubic Lagrangian is that it is defined in the 
$2$-dimensional space-time, while the original Lagrangian
(\ref{lag1}) and the quartic Lagrangian (\ref{lag4}) 
is defined for all space-time dimensions.
\end{enumerate}

Thus, we have constructed two classically equivalent Lagrangians $\mathcal{L}_4$ and 
$\mathcal{L}_3$ that appear to be quite different, with the explicit symmetries of one theory being implicit in the other. 

Note that although the cubic Lagrangian is classically identical to the $SU(2)$ sigma model, the quantum field theories are radically different, with the sigma model being asymptotically free and the cubic Lagrangian being non-asymptotically free \cite{nappi}. We consequently separately compute the beta function of both the quartic and cubic theories to check whether the classical equivalence that we have derived remains valid when the fields are quantized. 

Since -- unlike equivalent theories related by a duality transformation --  both the quartic and cubic Lagrangians have a common weak coupling sector, we can compare the quantum behaviour of the two theories using weak coupling perturbation theory.
 
\section{\normalsize \bf One-Loop Renormalization of 
$ \mathcal{L}_4 $ and $ \mathcal{L}_3 $} 

The background field method is used to obtain the one-loop 
$\beta$-function. We will obtain the result that both $ \mathcal{L}_4 $ and $ \mathcal{L}_3 $ are one-loop renormalizable, and with the same beta function, showing that the two theories are equivalent as quantum fields. 

Consider the quartic Lagrangian
\begin{eqnarray}
\mathcal{L}_4 = \frac{ 1}{ 2 g^2 } \mbox{\large[} \partial_{\mu} 
\alpha^i \partial_{\mu} \alpha^i + \mbox{\large(} \partial_{\mu}
\phi - \epsilon_{ij} \alpha^i \partial_{\mu} \alpha^j
\mbox{\large)}^2 \mbox{\large]} \, . \label{lag4b}
\end{eqnarray}
The qauntum field theory has both ultra-violet (UV) and infrared divergences.
To perform the one-loop calculation, we use a UV-regulated propagator 
by replacing
\begin{eqnarray}
\frac{1}{ p^2 } & \rightarrow & \frac{ e^{- p^2 / \Lambda^2 } }{p^2 } \nonumber
\end{eqnarray}
The infrared divergences are regulated by evaluating the Feynman diagrams in the dimension  
$ d = 2 + \epsilon $; we consequently have a dimesional coupling constant $g^2\Lambda^{-\epsilon}$.

The action is given by 
\begin{eqnarray}
S _4[ \phi ; \alpha^i ] = \frac{1}{2 g^2\Lambda^{-\epsilon}} \int d^{2 + \epsilon}
x \, \mbox{\large[} \partial_{\mu} \alpha^i \partial_{\mu} \alpha^i + 
 \mbox{\large(} \partial_{\mu} 
\phi - \epsilon_{ij} \alpha^i 
\partial_{\mu} \alpha^j \mbox{\large)}^2 \mbox{\large]} \, . 
\end{eqnarray}
Using the standard approach \cite{thooft}, we define the background fields by
\begin{eqnarray}
\phi & \rightarrow & \phi + \Phi \nonumber \\
\alpha^i & \rightarrow & \alpha^i + A^i \nonumber 
\end{eqnarray}
where $ \phi $, $ \alpha^i $ are the quantum fields and 
$ \Phi $, $ A^i $ are the background fields. The generating
functional for the quartic Lagrangian is then given by
\begin{eqnarray}
Z &=& \int D \phi D \alpha^i e^{-S_4 
[ \phi + \Phi ; \alpha^i + A^i ]}   \nonumber 
\end{eqnarray}

Let the contribution from the one-loop be denoted by $\Delta \mathcal{L}_4$. The one-loop calculation yields
\begin{eqnarray}
\mathcal{L}_4 + \Delta \mathcal{L}_4 =  
\frac{1}{2 g^2} \left[
\partial_\mu A^i \partial_\mu A^i \left( 1 + 
\frac{g^2 \hbar \Lambda^{\epsilon}}{\pi \epsilon} \right) + 
\mbox{\large(} \partial_{\mu}
\Phi - \epsilon_{ij} A^i \partial_{\mu}
A^j \mbox{\large)}^2  \left( 1 - 
\frac{g^2 \hbar \Lambda^{\epsilon}}{\pi \epsilon} \right) 
\right] 
\end{eqnarray}

The renormalized Lagrangian is defined at some momentum scale $\mu$ with renormalized coupling constant $g_R^2\mu^{-\epsilon}$. The one-loop renormalized Lagrangian is given by
\begin{eqnarray}
\mathcal{L}^{\mathrm{RN}}_4  [ \Phi_R , A_R,g_R ] = \mathcal{L}_4+\Delta \mathcal{L}_4 
\end{eqnarray}
where
\begin{eqnarray}
\Phi = Z_{\phi}^\frac{1}{2} \Phi_R ~~;~~A^i = Z_{A}^\frac{1}{2} A_R^i ~~;~~ g = (\frac{\Lambda}{\mu}) ^{\frac{\epsilon}{2}} g_RZ_g  
\end{eqnarray}
We have three renormalization constants, namely $Z_{\phi}, Z_{A}$ and $Z_{g}$ and only two constants from the one-loop calculation. To recover the original vertex we impose the condition
\begin{eqnarray}
Z_\Phi = Z_{A}^2 
\end{eqnarray}
We can then uniquely fix the two remaining renormalization constants and obtain
\begin{eqnarray}
Z_\Phi = 1 + \frac{ 2 g_R^2}{ \pi \epsilon }(\frac{\Lambda}{\mu}) ^\epsilon  ~~;~~
Z_{A} = 1 + \frac{g_R^2}{ \pi \epsilon}(\frac{\Lambda}{\mu}) ^\epsilon   ~~;~~
Z_{g} = 1 + \frac{3 g_R^2}{ 2 \pi \epsilon }(\frac{\Lambda}{\mu}) ^\epsilon  
\end{eqnarray}

To compute the $\beta$ funtion, we adopt Wilson's point of view that the bare coupling constant $g$ is a function of the UV-cutoff $\Lambda$, and varies in a manner so that the renormalized coupling constant $g_R$ is independent of the cutoff. The $\beta$ funtion is defined for the dimensionless coupling constant; we hence have
\begin{eqnarray}
\beta \equiv  \Lambda \frac{ \partial g}{ \partial \Lambda }
\end{eqnarray}
with
\begin{eqnarray}
\frac{ \partial g_R}{ \partial \Lambda }= 0
\end{eqnarray}
We hence have the $ \beta $ function given by
\begin{eqnarray}
\beta & = & \frac{ \epsilon}{2 } g_R (\frac{\Lambda}{\mu}) ^{\frac{\epsilon}{2}}  
+ \frac{3 g_R^3}{2 \pi } (\frac{\Lambda}{\mu}) ^\epsilon   + O(g_R^4)\\
& \rightarrow & \frac{3}{2 \pi } g_R^3  +O(\epsilon)\, .
\end{eqnarray}
Thus we have a positive beta function for small $g$ for the
theory, and hence the theory is not asymptotically free. 

A similar one-loop calculation for the cubic Lagrangian yields the
renormalized Lagrangian to be 
\begin{eqnarray}
\mathcal{L}^{\mathrm{RN}}_3  [ \Phi_R , A_R, g_R ] = \mathcal{L}_3+\Delta \mathcal{L}_3 
\end{eqnarray}
where
\begin{eqnarray}
Z_{\Phi} &=& 1 - \frac{ g_R^2}{\pi \epsilon }(\frac{\Lambda}{\mu}) ^\epsilon    \qquad \qquad ~;~\qquad Z_{g} \, = \, 
1 + \frac{ 3 g_{R}^{2}}{ 2 \pi \epsilon }(\frac{\Lambda}{\mu}) ^\epsilon    \label{zphig3} 
\end{eqnarray}
Note that while $Z_g$ remains unchanged in going from the quartic to 
the cubic case, $Z_{\Phi}$ is different. Hence we obtain the same $\beta$ function for the cubic Lagrangian, establishing its one-loop equivalence to the quartic theory.
 
\

\section{ \normalsize \bf Conclusions }
 
\

\indent
We have analyzed the Heisenberg 
sigma-model Lagrangian. The Lagrangian needs to be regularized, and we 
found that with a natural regulator, the Lagrangian 
has a quartic interaction. Moreover, it can be written 
as a cubic and a quartic Lagrangian with very different manifest symmetries. 

One particularly interesting feature about this theory is when 
the dimension of space-time is two. Here, we can reformulate the real quartic Lagrangian as an equivalent theory with an imaginary cubic interaction. The two equivalent theories are not related by a duality transformation in that the coupling constants for the two theories are not inversely related. By computing the one-loop $\beta$ function we explicitly demostrated the one-loop quantum equivalence of the two theories. The equivalence of the cubic and quartic Lagrangians is to our knowledge the first instance of two apparently different two-dimensional bosonic theories being equivalent as quantum field theories. 

We also found a simple explanation as to why the classical equivalence of the SU(2) cubic (pseudo-chiral) Lagrangian with the principal SU(2) chiral model breaks down: the cubic Lagrangian results from a sigma model based on the non-compact Heisenberg group. Under renormalization this difference in the target space causes the classical equivalence to break down.

{\bf Acknowledgement}

We have benefitted from many fruitful discussions with Rajesh Parwani.

\end{document}